\documentclass[conference]{IEEEtran}

\usepackage{graphicx}
\usepackage[caption=false, font=footnotesize]{subfig}
\usepackage{amsmath,amssymb,amsthm}
\usepackage{mathabx}
\usepackage{dsfont}
\usepackage{bm}
\usepackage{rotating}
\usepackage[lined,ruled]{algorithm2e}
\usepackage{multirow}
\usepackage{rotating}
\usepackage{multicol}
\usepackage{float}
\usepackage{latexsym}
\usepackage[pdftex,colorlinks=true,linkcolor=blue,citecolor=blue,urlcolor=blue,bookmarks=false,pdfpagemode=None]{hyperref}
\usepackage{url}
\usepackage{verbatim}
\usepackage{fancyhdr}
\usepackage{setspace}
\usepackage{paralist}
\usepackage{rotating}
\usepackage{lineno}
\usepackage[top=1in, bottom=1in, left=1in, right=1in]{geometry}
\usepackage{lscape}
\usepackage{dcolumn}
\usepackage{color, colortbl}
\usepackage{hhline}
\usepackage{enumitem}

\makeatletter
\newcommand*\bigcdot{\mathpalette\bigcdot@{.5}}
\newcommand*\bigcdot@[2]{\mathbin{\vcenter{\hbox{\scalebox{#2}{$\m@th#1\bullet$}}}}}
\makeatother

\newcommand{\A}{\bm{A}}
\newcommand{\N}{\mathcal{N}}
\newcommand{\M}{\mathcal{M}}

\definecolor{LightGreen}{rgb}{0.85,1,0.85}
\definecolor{LightPink}{rgb}{1,0.85,0.85}

\begin{document}
\title{\fontsize{21}{26}\selectfont Streaming Graph Challenge: Stochastic Block Partition}

% author names and affiliations
\author{\IEEEauthorblockN{Edward Kao,
Vijay Gadepally,
Michael Hurley,
Michael Jones,
Jeremy Kepner,
Sanjeev Mohindra, \\
Paul Monticciolo,
Albert Reuther,
Siddharth Samsi,
William Song,
Diane Staheli,
Steven Smith \\
\IEEEauthorblockA{MIT Lincoln Laboratory, Lexington, MA}}}

% make the title area
\maketitle
\begin{abstract}
An important objective for analyzing real-world graphs is to achieve scalable performance on large, streaming graphs. A challenging and relevant example is the graph partition problem. As a combinatorial problem, graph partition is NP-hard, but existing relaxation methods provide reasonable approximate solutions that can be scaled for large graphs. Competitive benchmarks and challenges have proven to be an effective means to advance state-of-the-art performance and foster community collaboration. This paper describes a graph partition challenge with a baseline partition algorithm of sub-quadratic complexity. The algorithm employs rigorous Bayesian inferential methods based on a statistical model that captures characteristics of the real-world graphs. This strong foundation enables the algorithm to address limitations of well-known graph partition approaches such as modularity maximization. This paper describes various aspects of the challenge including: (1) the data sets and streaming graph generator, (2) the baseline partition algorithm with pseudocode, (3) an argument for the correctness of parallelizing the Bayesian inference, (4) different parallel computation strategies such as node-based parallelism and matrix-based parallelism, (5) evaluation metrics for partition correctness and computational requirements, (6) preliminary timing of a Python-based demonstration code and the open source C++ code, and (7) considerations for partitioning the graph in streaming fashion. Data sets and source code for the algorithm as well as metrics, with detailed documentation are available at \href{http://GraphChallenge.org}{GraphChallenge.org}. \let\thefootnote\relax\footnotetext{*This material is based upon work supported by the Defense Advanced Research Projects Agency under Air Force Contract No. FA8721-05-C-0002. Any opinions, findings and conclusions or recommendations expressed in this material are those of the author(s) and do not necessarily reflect the views of the Department of Defense.}
\end{abstract}
% no keywords

\IEEEpeerreviewmaketitle

\section{Introduction}
In the era of big data, analysis and algorithms often need to scale up to large data sets for real-world applications. With the rise of social media and network data, algorithms on graphs face the same challenge. Competitive benchmarks and challenges have proven to be an effective means to advance state-of-the-art performance and foster community collaboration. Previous benchmarks such as Graph500 \cite{murphy2010} and the Pagerank Pipeline \cite{dreher2016} are examples of such, targeting analysis of large graphs and focusing on problems with sub-quadratic complexity, such as search, path-finding, and PageRank computation. However, some analyses on graphs with valuable applications are NP-hard. The graph partition and the graph isomorphism (i.e.\ matching) problems are well-known examples. Although these problems are NP-hard, existing relaxation methods provide good approximate solutions that can be scaled to large graphs \cite{jin2016, kanezashi2016}, especially with the aid of high performance computing hardware platform such as massively parallel CPUs and GPUs. For example, the 10th DIMACS Implementation Challenge \cite{bader2013} resulted in substantial participation in the graph partition problem, mostly with solutions based on modularity maximization. To promote algorithmic and computational advancement in these two important areas of graph analysis, our team has implemented a challenge for graph isomorphism \cite{Samsi2017} and graph partition at \href{http://GraphChallenge.org}{GraphChallenge.org}. This paper describes the graph partition challenge with a recommended baseline partition algorithm of sub-quadratic complexity. Furthermore, the algorithm employs rigorous Bayesian inferential methods based on the stochstic blockmodels that capture characteristics of the real-world graphs. Participants are welcome to submit solutions based on other partition algorithms as long as knowledge on the true number of communities (i.e.\ blocks) is not assumed. All entries should be submitted with performance evaluation on the challenge data sets using the metrics described in Section \ref{sec:metrics}.

Graph partition, also known as community detection and graph clustering, is an important problem with many real-world applications. The objective of graph partition is to discover the distinct community structure of the graph, specifically the community membership for each node in the graph. The partition gives much insight to the interactions and relationships between the nodes and enables detection of nodes belonging to certain communities of interest. Much prior work has been done in the problem space of graph partition, with a comprehensive survey in \cite{Fortunato2010}. The most well-known algorithm is probably the spectral method by \cite{Newman2006} where partition is done through the eigenspectrum of the modularity matrix. Most of the existing partition algorithms work through the principle of graph modularity where the graph is partitioned into communities (i.e.\ modules) that have much stronger interactions within them than between them. Typically, partitioning is done by maximizing the graph modularity \cite{Newman2006modularity}. \cite{Mucha2010} extends the concept of modularity for time-dependent, multiscale, and multiplex graphs. Modularity maximization is an intuitive and convenient approach, but has inherent challenges such as resolution limit on the size of the detectable communities \cite{lancichinetti2011}, degeneracies in the objective function, and difficulty in identifying the optimal number of communities \cite{good2010}. 

To address these challenges, recent works perform graph partition through membership estimation based on generative statistical models. For example, \cite{Ball2011, peixoto2014, peixoto2013, peixoto2012} estimate community memberships using the degree corrected stochastic blockmodels \cite{karrer2011}, and \cite{Huang2013} proposes a mixed-memberships estimation procedure by applying tensor methods to the mixed-membership stochastic blockmodels \cite{Airoldi2008}. The baseline partition algorithm for this challenge is based on \cite{peixoto2014, peixoto2013, peixoto2012}, because of its rigorous statistical foundation and sub-quadratic computational requirement. Under this approach, each community is represented as a ``block'' in the model. Going forward, this paper will use the term ``block'' as the nomenclature for a community or a graph cluster.

When some nodes in the graph have known memberships a priori, these nodes can serve as ``cues'' in the graph partition problem. \cite{Smith2014} is an example of such using random walks on graph. This challenge will focus on the graph partition problem where such cues are not available.

In many real-world applications, graph data arrives in streaming fashion over time or stages of sampling \cite{ahmed2014}. This challenge addresses this aspect by providing streaming graph data sets and recommending a baseline partition algorithm that is suitable for streaming graphs under the Bayesian inference paradigm. 

This paper describes the graph partition challenge in detail, beginning with Section \ref{sec:datasets} on the data sets and streaming graph generator. Section \ref{sec:partition_baseline_algorithm} describes the baseline partition algorithm, including pseudocode on the core Bayesian updates. Section \ref{sec:parallelComputation} focuses on the parallel computation of the baseline algorithm, argues for the correctness of parallelizing the Bayesian updates, then proposes parallel computation strategies such as node-based parallelism and matrix-based parallelism. Section \ref{sec:metrics} describes the evaluation metrics for both partition correctness and computational requirements, including a preliminary timing of a Python-based demonstration code and the open source C++ code by Tiago Peixoto \cite{graph-tool_github}. Considerations for partitioning the graph in streaming fashion are given throughout the paper. 

\section{Data Sets}
\label{sec:datasets}
The data sets for this challenge consist of graphs of varying sizes and characteristics. Denote a graph $\bm{G} = (\bm{\mathcal{V}}, \bm{\mathcal{E}})$, with the set $\bm{\mathcal{V}}$ of $N$ nodes and the set $\bm{\mathcal{E}}$ of $E$ edges. The edges, represented by a $N \times N$ adjacency matrix $\A$, can be either directed or undirected, binary or weighted. Specifically, $A_{ij}$ is the weight of the edge from node $i$ to node $j$. A undirected graph will have a symmetric adjacency matrix.

\begin{figure}[h] 
  \vspace{-0.5cm}
  \centering
  \subfloat[baseline]{%
       \includegraphics[width=0.5\linewidth]{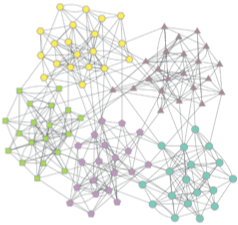}}
    \label{1a}\hfill
  \subfloat[increased block overlap]{%
        \includegraphics[width=0.5\linewidth]{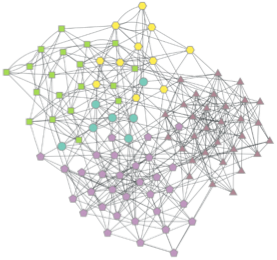}}
    \label{1b}\\
  \subfloat[higher block size variation]{%
        \includegraphics[width=0.5\linewidth]{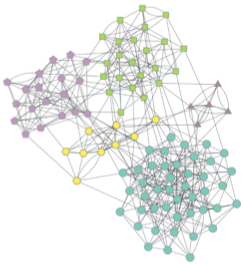}}
    \label{1c}\hfill
  \subfloat[more high degree nodes]{%
        \includegraphics[width=0.5\linewidth]{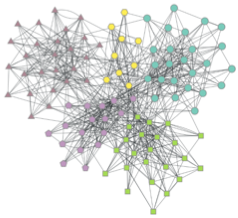}}
     \label{1d} 
  \caption{Generated graphs with varying characteristics. Nodes are colored and shaped according to their true block assignments. Graphs are typically much larger. Small graphs are shown here for the purpose of demonstration. For simplicity and clarity, the edge directions (i.e.\ arrows) are not displayed.}
  \label{fig:graphsWithVaryingProperties} 
\end{figure}

In order to evaluate the partition algorithm implementation on graphs with a wide range of realistic characteristics, graphs are generated according to a truth partition $\bm{b^}\dagger$ of $B^\dagger$ blocks (i.e.\ clusters), based on the degree-corrected stochastic blockmodels by Karrer and Newman in \cite{karrer2011}. Under this generative model, each edge, $A_{ij}$, is drawn from a Poisson distribution of rate $\lambda_{ij}$ governed by the equations below: 
\begin{align} 
	A_{ij} &\sim \text{Poisson}(\lambda_{ij}) \\
	\lambda_{ij} &= \theta_i \theta_j \Omega_{b_i b_j}
\end{align}

\noindent where $\theta_i$ is a correction term that adjusts node $i$'s expected degree, $\Omega_{b_i b_j}$ the strength of interaction between block $b_i$ and $b_j$, and $b_i$ the block assignment for node $i$. The degree-corrected stochastic blockmodels enable the generation of graphs with characteristics and variations consistent with real-world graphs. The degree correction term for each node can be drawn from a Power-Law distribution with an exponent between $-3$ and $-2$ to capture the degree distribution of realistic, scale-free graphs \cite{barabasi2009}. The block interaction matrix $\bm{\Omega}$ specifies the strength of within- and between-block (i.e.\ community) interactions. Stronger between-block interactions will increase the block overlap, making the block partition task more difficult. Lastly, the block assignment for each node (i.e.\ the truth partition $\bm{b^}\dagger$) can be drawn from a multinomial distribution with a Dirichlet prior that determines the amount of variation in size between the blocks. Figure~\ref{fig:graphsWithVaryingProperties} shows generated graphs of various characteristics by adjusting the parameters of the generator. These parameters server as ``knobs'' that can be dialed to capture a rich set of characteristics for realism and also for adjusting the difficulty of the block partition task. 

Real-world graphs will also be included in the data sets. Since the truth partition is not available in most real-world graphs, generated graphs with truth will be embedded with the real-world graphs. While the entire graph will be partitioned, evaluation on the correctness of the partition will be done only on the generated part of the hybrid graph. Embedding will be done by adding edges between nodes in the real-world graph and the generated graph, with a relatively small probability proportional to the product of both node degrees.  

\begin{figure}[h] 
  \vspace{0cm}
  \centering
  \subfloat[streaming graph as edges emerge]{%
       \includegraphics[width=1\linewidth]{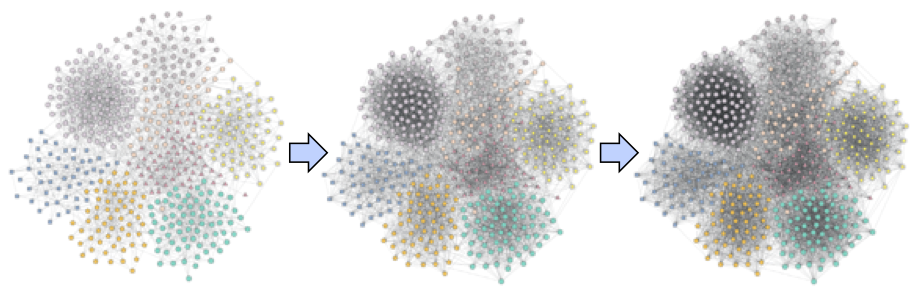}}
    \label{2a}\\
  \subfloat[streaming graph with snowball sampling]{%
        \includegraphics[width=1\linewidth]{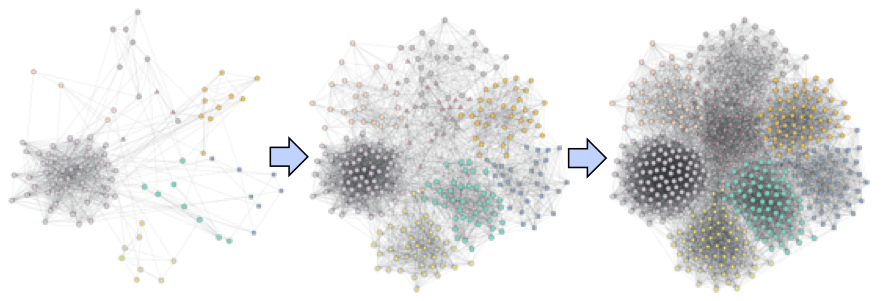}}
    \label{2b}
  \caption{Streaming graphs generated in two ways: (a) as edges emerge over time and (b) as the graph is explored from starting point(s).}
  \label{fig:streamingGraphs} 
\end{figure}

In real-world applications, graph data often arrives in streaming fashion, where parts of the input graph become available at different stages. This happens as interactions and relationships take place and are observed over time, or as data is collected incrementally by exploring the graph from starting points (e.g.\ breadth first search and snowball sampling) \cite{ahmed2014}. Streaming graph data sets in this challenge are generated in both ways, as demonstrated in Figure \ref{fig:streamingGraphs}. The partition algorithm should process the streaming graph at each stage and ingest the next stage upon completion of the current stage. Performance evaluated using the metrics in Section \ref{sec:metrics} should be reported at each stage of the processing. For efficiency, it is recommended that the partition algorithm leverages partitions from the previous stage(s) to speed up processing at the current stage. The baseline partition algorithm for this challenge is a natural fit for streaming processing, as discussed in Section \ref{sec:partition_baseline_algorithm}.

\section{Baseline Algorithm}
\label{sec:partition_baseline_algorithm}
This section described the recommended baseline partition algorithm, although participants are welcome to submit solutions based on other partition algorithms as long as knowledge on the true number of blocks is not assumed. 

The baseline graph partition algorithm for this challenge, chosen for its rigorous statistical foundation and sub-quadratic, $O(E \log^2 E)$, computational requirement, is developed by Tiago Peixoto in \cite{peixoto2014, peixoto2013, peixoto2012} based on the degree-corrected stochastic blockmodels by Karrer and Newman in \cite{karrer2011}. Given the input graph, the algorithm partitions the nodes into $B$ blocks (i.e.\ clusters or communities), by updating the nodal block assignment represented by vector $\bm{b}$ of $N$ elements where $b_i \in \{1, 2, ..., B\}$, and the inter-block and intra-block edge count matrix (typically sparse in a large graph) represented by $\bm{M}$ of size $B \times B$, where each element  $M_{ij}$ represents the number or the total weight of edges going from block $i$ to block $j$. The diagonal elements represent the edge counts within each block. For conciseness, this matrix will be referred to as the inter-block edge count matrix going forward. The goal of the algorithm is to recover the truth partition $\bm{b^}\dagger$ of $B^\dagger$ blocks (i.e.\ clusters).

The algorithm performs a Fibonacci search (i.e.\ golden section search) \cite{numerical1982} through different numbers of blocks $B$ and attempts to find the minimum description length partition. The best overall partition $\bm{b}^*$ with the optimal number of block $B^*$ minimize the total description length of the model and the observed graph (i.e.\ entropy of the fitted model). To avoid being trapped in local minima, the algorithm starts with each node in its own block (i.e.\ $B=N$) and the blocks are merged at each step of the Fibonacci search, followed by iterative Monte Carlo Markov Chain (MCMC) updates on the block assignment for each node to find the best partition for the current number of blocks. The block-merge moves and the nodal updates are both governed by the same underlying log posterior probability of the partition given the observed graph:

\begin{equation}
\label{eq:posterior}
     p(\bm{b} | G) \;\; \propto \; \sum_{t_1, t_2} {M_{t_1 t_2} \log\left(\frac{M_{t_1 t_2}}{d_{t_1,\rm out} d_{t_2,\rm in}}\right)}
\end{equation}

\noindent The log posterior probability is a summation over all pairs of blocks $t_1$ and $t_2$ where $d_{t_1,\rm out}$ is the total out-degree for block $t_1$ and $d_{t_2, \rm in}$ is the total in-degree for block $t_2$. Note that in computing the posterior probabilities on the block assignments, the sufficient statistics for the entire graph is only the inter-block edge counts, giving much computational advantage for this algorithm.
Another nice property of the log posterior probability is that it is also the negative entropy of the fitted model. Therefore, maximizing the posterior probability of the partition also minimizes the overall entropy, fitting nicely into the minimum description length framework. The block-merge moves and the nodal block assignment updates are described in detail next, starting with the nodal updates.

\subsection{Nodal Block Assignment Updates}
The nodal updates are performed using the MCMC, specifically with Gibbs sampling and the Metropolis-Hastings algorithm since the partition posterior distribution in Equation \ref{eq:posterior} does not have a closed-form and is best sampled one node at a time. At each MCMC iteration, the block assignment of each node $i$ is updated conditional on the assignments of the other nodes according to the conditional posterior distribution: $p(b_i | \bm{b}_{-i}, G)$. Specifically, the block assignment $b_i$ for each node $i$ is updated based on the edges to its neighbors, $\bm{A}_{i \N_i}$ and $\bm{A}_{\N_i i}$, the assignments of its neighbors, $\bm{b}_{\N_i}$, and the inter-block edge count, $\bm{M}$. For each node $i$, the update begins by proposing a new block assignment. To increase exploration, a block is randomly chosen as the proposal with some predefined probability. Otherwise, the proposal will be chosen from the block assignments of nodes nearby to $i$. The new proposal will be considered for acceptance according to how much it changes the log posterior probability. The acceptance probability is adjusted by the Hastings correction, which accounts for potential asymmetry in the directions of the proposal to achieve the important detailed balance condition that ensures the correct convergence of the MCMC. Algorithm \ref{alg:NodalUpdate} in Appendix \hyperref[sec:appendixA]{A} is a detailed description of the block assignment update at each node, using some additional notations: $d_{t, \rm in} = \sum_k{M_{kt}}$ is the number of edges into block $t$,  $d_{t, \rm out} = \sum_k{M_{tk}}$ the number of edges out of block $t$, $d_t = d_{t, \rm  in} + d_{t, \rm out}$ the number of edges into and out of block $t$, $K_{it}$ the number of edges between nodes $i$ and block $t$, and $\beta$ is the update rate that controls the balance between exploration and exploitation. The block assignments are updated for each node iteratively until convergence when the improvement in the log posterior probability falls below a threshold. 

\subsection{Block-Merge Moves}
The block-merge moves work in almost identical ways as the nodal updates described in Algorithm \ref{alg:NodalUpdate} in Appendix \hyperref[sec:appendixA]{A}, except that it takes place at the block level. Specifically, a block-merge move proposes to reassign all the nodes belonging to the current block $i$ to a proposed block $s$. In other words, it is like applying Algorithm \ref{alg:NodalUpdate} on the block graph where each node represents the entire block (i.e.\ all the nodes belonging to that block) and each edge represents the number of edges between the two blocks. Another difference is that the block-merges are done in a greedy manner to maximize the log posterior probability, instead of through MCMC. Therefore, the Hastings correction computation step and the proposal acceptance step are not needed. Instead, the best merge move over some number of proposals is computed for each block according to the change in the log posterior probability, and the top merges are carried out to result in the number of blocks targeted by the Fibonacci search.

\subsection{Put It All Together}
Overall, the algorithms shifts back and forth between the block-merge moves and the MCMC nodal updates, to find the optimal number of blocks $B^*$ with the resulting partition $\bm{b}^*$. Optimality is defined as having the minimum overall description length, $H$, of the model and the observed graph given the model:
 
\begin{equation}
\label{eq:totalEntropy}
 	H = E \; h\left(\frac{B^2}{E}\right)+N\log B- \sum_{r,s}{M_{rs} \log\left(\frac{M_{rs}}{d_{r,\rm out} d_{s, \rm in}}\right)}
\end{equation}
 
\noindent where the function $h(x) = (1+x) \log(1+x) - x\log(x)$. The number of blocks may be reduced at a fixed rated (e.g.\ $50\%$) at each block-merge phase until the Fibonacci 3-point bracket is established. At any given stage of the search for optimal number of blocks, the past partition with the closest and higher number of blocks is used to begin the block-merge moves, followed by the MCMC nodal updates, to find the best partition at the targeted number of blocks. Figure \ref{fig:partitionResult} shows the partition at selected stages of the algorithm on a $500$ node graph:
\begin{figure}[h] 
  \vspace{-0.5cm}
  \centering
  \subfloat[$250$ blocks]{%
       \includegraphics[width=0.5\linewidth]{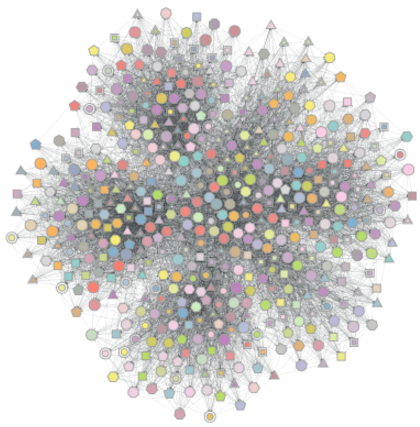}}
    \label{3a}\hfill
  \subfloat[$32$ blocks]{%
        \includegraphics[width=0.5\linewidth]{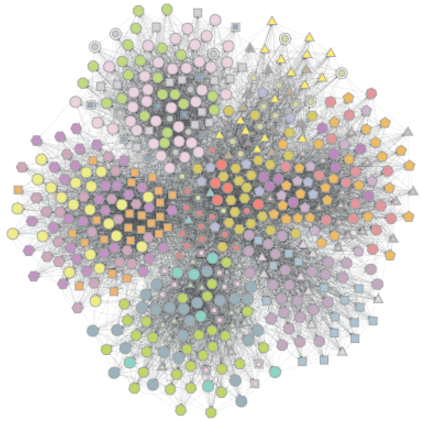}}
    \label{3b}\\
  \subfloat[$8$ blocks]{%
        \includegraphics[width=0.5\linewidth]{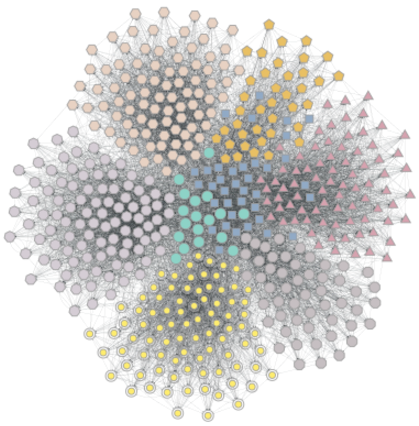}}
    \label{3c}\hfill
  \subfloat[$4$ blocks]{%
        \includegraphics[width=0.5\linewidth]{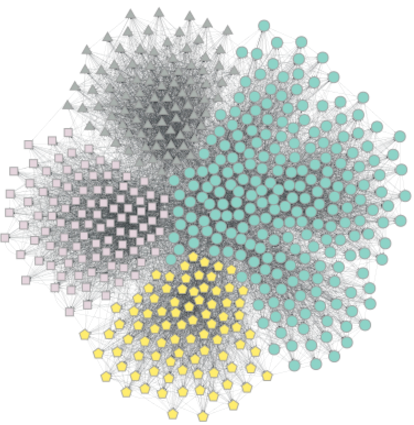}}
     \label{3d} 
  \caption{Partitions at selected stages of the algorithm, with the nodes colored and shaped according to their block assignments. The algorithm begins with too many blocks (i.e.\ over partition) and performs block-merges and nodal updates as it searches for the optimal partition. The Fibonacci search eventually converges to the partition with the optimal number of blocks, which is shown in (c) with $8$ blocks.}
  \label{fig:partitionResult} 
\end{figure}

The algorithm description in this section is for directed graphs. Very minor modifications can be applied for undirected graphs that have no impact on the computational requirement. These minor differences are documented in Peixoto's papers \cite{peixoto2014, peixoto2013, peixoto2012}.

Advantageously, the baseline partition algorithm with its rigorous statistical foundation, is ideal for processing streaming graphs. Good partitions found on the graph at a previous streaming stage are samples on the posterior distribution of the partition, which can be used as starting partitions for the graph at the current stage. This has the natural Bayesian interpretation of the posterior distribution from a previous state serving as the prior distribution on the current state, as additional data on the graph arrives.

\section{Parallel Computation Strategies}
\label{sec:parallelComputation}

Significant speed up of the baseline partition algorithm is the primary focus of this graph challenge, and is necessary for computation on large graphs. Since the same core computation, described in Algorithm \ref{alg:NodalUpdate} in Appendix \hyperref[sec:appendixA]{A}, is repeated for each block and each node, parallelizing this core computation across the blocks and nodes provides a way to speed up the computation potentially by the order of the number of processors available. This section first discusses the correctness in parallelizing the MCMC updates. It then examines some of the parallel computation schemes for the baseline algorithm, with their respective advantages and requirements.

\subsection{Correctness of Parallel MCMC Updates}

The block-merge moves are readily parallelizable, since each of the potential merge move is evaluated based on the previous partition and the best merges are carried out. However, the nodal block assignment updates are not so straight forward, since it relies on MCMC through Gibbs sampling which is by nature a sequential algorithm where each node is updated one at a time. Parallelizing MCMC updates is an area of rising interest, with the increasing demand to perform Bayesian inference on large data sets. Running the baseline partition algorithm on large graphs is a perfect example of this need. Very recently, researchers have proposed to use asynchronous Gibbs sampling as a way to parallelize MCMC updates \cite{terenin2015, de2016}. In asynchronous Gibbs sampling, the parameters are updated in parallel and asynchronous fashion without any dependency constraint. In \cite{de2016}, a proof is given to show that when the parameters in the MCMC sparsely influence one another (i.e.\ the Dobrushin's condition), asynchronous Gibbs is able to converge quickly to the correct distribution. It is difficult to show analytically that the MCMC nodal updates here satisfy the Dobrushin's condition. However, since the graph is typically quite sparse, the block assignment on each node influences one another sparsely. This gives intuition on the adequacy of parallel MCMC updates for the baseline partition algorithm. In fact, parallel MCMC updates based on one-iteration-old block assignments have shown to result in equally good partitions compared to the sequential updates, based on the quantitative metrics in Section \ref{sec:correctnessMetrics}, for the preliminary tests we conducted so far. 

\subsection{Parallel Updates on Nodes and Blocks}

An intuitive and straight-forward parallel computation scheme is to evaluate each block-merge and update each nodal block assignment (i.e.\ Algorithm \ref{alg:NodalUpdate} in Appendix \hyperref[sec:appendixA]{A}) in distributed fashion across multiple processors. The block-merge evaluation is readily parallelizable since the computation is based on the previous partition. The MCMC nodal updates can be parallelized using the one-iteration-old block assignments, essentially approximating the true conditional posterior distribution with: $p(b_i | \bm{b}^-_{-i}, G)$. The conditional block assignments, $\bm{b}^-_{-i}$, may be more ``fresh'' if asynchronous Gibbs sampling is used so that some newly updated assignments may become available to be used for updates on later nodes. In any case, once all the nodes have been updated in the current iteration, all the new block assignments are gathered and their modifications on the inter-block edge count matrix aggregated (this can also be done in parallel). These new block assignments and the new inter-block edge count matrix are then available for the next iteration of MCMC updates.   

\subsection{Batch Updates Using Matrix Operations}

Given an efficient parallelized implementation of large-scale matrix operations, one may consider carrying out Algorithm \ref{alg:NodalUpdate} as much as possible with batch computation using matrix operations \cite{kepner2011}. Such matrix operations in practice perform parallel computation across all nodes simultaneously. 

Under this computation paradigm, the block assignments are represented as a sparse $N \times B$ binary matrix $\bm{\Gamma}$, where each row $\bm{\pi}_{i \bigcdot}$ is an indicator vector with a value of one at the block it is assigned to and zeros everywhere else. This representation results in simple matrix products for the inter-block edge counts:
\begin{equation}
\bm{M} = \bm{\Gamma}^T\!\!\bm{A}\bm{\Gamma}
\end{equation}
The contributions of node $i$ of block assignment $r$ to the inter-block edge count matrix row $r$ and column $r$ are:  
\begin{align} 
	{\Delta \bm{M}}_{{\rm row},i\bigcdot}  &= \bm{A}_{i\bigcdot}\bm{\Gamma} \\
	{\Delta \bm{M}}^+_{{\rm col},i\bigcdot} &= \A_{\bigcdot i}^T\bm{\Gamma}
\end{align}
These contributions are needed for computing the acceptance probabilities of the nodal block assignment proposals, which makes up a large part of the overall computation requirement.

Algorithm \ref{alg:batchAssignmentUpdate} in Appendix \hyperref[sec:appendixB]{B} is a batch implementation of the nodal updates described in Algorithm \ref{alg:NodalUpdate}. The inter-block edge counts under each of the $N$ proposal are represented using a 3D matrix $\bm{\M}$ of size $N \times B \times B$. For clarity, computations of the acceptance probabilities involving the inter-block edge counts and degrees are specified using tensor notation. Note that much of these computations may be avoided with clever implementations. For example:
\begin{itemize}
\item If the proposed block assignment for a node is the same as its previous assignment, its acceptance probability does not need to be computed. 
\item New proposals only change two rows and columns of the inter-block edge count matrix, corresponding to moving the counts from the old block to the new block, so most of the entries in $\bm{\M}$ are simply copies of $\bm{M}^-$.  
\item The inter-block edge count matrix should be sparse, especially when there is a large number of communities, since most communities do not interact with one another. This gives additional opportunity for speeding up operations on this matrix.
\item Similarly, each node is likely to connect with only a few different communities (i.e.\ blocks). Therefore, changes by each nodal proposal on the inter-block edge count matrix will only involve a few selected rows and columns. Limiting the computation of change in log posterior, $\Delta S$, to these rows and columns may result in significant computation speedup.
\end{itemize}

\section{Metrics}
\label{sec:metrics}

An essential part of this graph challenge is a canonical set of metrics for comprehensive evaluation of the partition algorithm implementation by each participating team. The evaluation should report both the correctness of the partitions produced, as well as the computational requirements, efficiency, and complexity of the implementations. For streaming graphs, evaluation should be done at each stage of the streaming processing, for example, the length of time it took for the algorithm to finish processing the graph after the first two parts of the graph become available, and the correctness of the output partition on the available parts so far. Efficient implementations of the partition algorithm leverage partitions from previous stages of the streaming graph to ``jump start'' the partition at the current stage.

\subsection{Correctness Metrics}
\label{sec:correctnessMetrics}
The true partition of the graph is available in this challenge, since the graph is generated with a stochastic block structure, as described in Section \ref{sec:datasets}. Therefore, correctness of the output partition by the algorithm implementation can be evaluated against the true partition. On the hybrid graphs where a generated graph is embedded within a real-world graph with no available true partition, correctness is only evaluated on the generated part. 

Evaluation of the output partition (i.e.\ clustering) against the true partition is well established in existing literature and a good overview can be found in \cite{meila2007}. Widely-adopted metrics fall under three general categories: unit counting, pair-wise counting, and information theoretic metrics. The challenge in this paper adopts all of them for comprehensiveness and recommends the pairwise precision-recall as the primary correctness metric for its holistic evaluation and intuitive interpretation. Computation of the correctness metrics described in this section are implemented in Python and shared as a resource for the participants at \href{http://GraphChallenge.org}{GraphChallenge.org}. Table \ref{tab:contingency} provides a simple example to demonstrate each metric, where each cell in row $i$ and column $j$ is the count of nodes belonging to truth block $i$ and reported in output block $j$. 

\begin {table}[h]
\caption{Contingency table of true vs. output partition}
\label{tab:contingency} 
\begin{center}
\vspace{-0.5cm}
\begin{tabular}{|c|c|c|c|c}
\hhline{~|*{3}{-}} 
\multicolumn{1}{c|}{}&Output A&Output B&Output C&\multicolumn{1}{c}{Total}\\    \hhline{*{4}{-}} 
Truth A & $\cellcolor{LightGreen} 30$ & $\cellcolor{LightPink}2$ & $0$ & $32$\\     \hhline{*{4}{-}} 
Truth B & $\cellcolor{LightPink}1$ & $\cellcolor{LightGreen}20$ & $\cellcolor{LightPink}3$ & $24$\\     \hhline{*{4}{-}} 
\multicolumn{1}{c}{Total} & \multicolumn{1}{c}{$31$} & \multicolumn{1}{c}{$22$} & \multicolumn{1}{c}{$3$} & \multicolumn{1}{c}{$56$}\\
\end{tabular}
\end{center}
\end {table}

\noindent In this example, the nodes are divided into two blocks in the true partition, but divided into three blocks in the output partition. Therefore, this is an example of over-clustering (i.e.\ too many blocks). The diagonal cells shaded in green here represent the nodes that are correctly partitioned whereas the off-diagonal cells shaded in pink represent the nodes with some kind of partition error.

\subsubsection{Unit Counting Metrics}
The most intuitive metric is perhaps the overall accuracy, specifically the percentage of nodes correctly partitioned. This is simply the fraction of the total count that belong to the diagonal entries of the contingency table after the truth blocks and the output blocks have been optimally associated to maximize the diagonal entries, typically using a linear assignment algorithm \cite{kuhn1955}. In this example, the overall accuracy is simply $50/66=89\%$. While this one single number provides an intuitive overall score, it does not account for the types and distribution of errors. For example, truth block B in Table \ref{tab:contingency} has three nodes incorrectly split into output block C. If instead, these three nodes were split one-by-one into output block C, D, and E, a worse case of over-clustering would have taken place. The overly simplified accuracy cannot make this differentiation. 

A way to capture more details on the types and distribution of errors is to report block-wise precision-recall. Block-wise precision is the fraction of correctly identified nodes for each output block (e.g.\ $\text{Precision(Output A)} = 30/31$) and the block-wise recall is the fraction of correctly identified nodes for each truth block (e.g.\ $\text{Recall(Truth B)} = 20/24$). The block-wise precision-recall present a intuitive score for each of the truth and output blocks, and can be useful for diagnosing the block-level behavior of the implementation. However, it does not provide a global measure on correctness.

\subsubsection{Pairwise Counting Metrics}
Measuring the level of agreement between the truth and the output partition by considering every pair of nodes has a long history within the clustering community \cite{rand1971, hubert1985}. The basic idea is simple, by considering every pair of nodes which belongs to one of the following four categories: (1) in the same truth block and the same output block, (2) in different truth blocks and different output blocks, (3) in the same truth block but different output blocks, and (4) in different truth blocks but the same output block. Category (1) and (2) are the cases of agreements between the truth and the output partition, whereas categories (3) and (4) indicate disagreements. An intuitive overall score on the level of agreement is the fraction of all pairs belonging to category (1) and (2), known as the Rand index \cite{rand1971}. \cite{ hubert1985} proposes the adjusted Rand index with a correction to account for the expected value of the index by random chance, to provide a fairer metric across different data sets. Categories (4) and (3) can be interpreted as type I (i.e.\ false positives) and type II (i.e.\ false negative) errors, if one considers a ``positive'' case to be where the pair belongs to the same block. The pairwise precision-recall metrics \cite{banerjee2005} can be computed as:

\begin{equation}
\text{Pairwise-precision} = \frac{\# \text{Category } 1}{\# \text{Category } 1 + \# \text{Category } 4 }
\end{equation}

\begin{equation}
\text{Pairwise-recall} = \frac{\# \text{Category } 1}{\# \text{Category } 1 + \# \text{Category } 3 }
\end{equation}

\noindent Pairwise-precision considers all the pairs reported as belonging to the same output block and measures the fraction of them being correct, whereas pairwise-recall considers all the pairs belonging to the same truth block and measures the fraction of them reported as belonging to the same output block. In the example of Table \ref{tab:contingency}, the pairwise-precision is about $90\%$ and the pairwise-recall about $81\%$, which indicates this to be a case of over-clustering with more Type II errors. Although pairwise counting is somewhat arbitrary, it does present holistic and intuitive measures on the overall level of agreement between the output and  the true partition. For the challenge, the pairwise precision-recall will serve as the primary metrics for evaluating correctness of the output partition.

\subsubsection{Information Theoretic Metrics}
In recent years, holistic and rigorous metrics have been proposed based on information theory, for evaluating partitions and clusterings \cite{meila2007, holt2010}. Specifically, these metrics are based on the information content of the partitions measured in Shannon entropy. Naturally, information theoretic precision-recall metrics can be computed as:

\begin{equation}
\text{Information-precision} = \frac{I(T;O)}{H(O)}
\end{equation}

\begin{equation}
\text{Information-recall} =  \frac{I(T;O)}{H(T)}
\end{equation}

\noindent where $I(T;O)$ is the mutual information between truth partition $T$ and the output partition $O$, and $H(O)$ is the entropy (i.e.\ information content) of the output partition. Using the information theoretic measures, precision is defined as the fraction of the output partition information that is true, and recall is defined as the fraction of the truth partition information captured by the output partition.  In the example of Table \ref{tab:contingency}, the information theoretic precision is about $57\%$ and recall about $71\%$. The precision is lower than the recall because of the extra block in the output partition introducing information content that does not correspond to the truth. The information theoretic precision-recall provide a rigorous and comprehensive measure of the correctness of the output partition. However, the information theoretic quantities may not be as intuitive to some and the metrics tend to be harsh, as even a small number of errors often lower the metrics significantly. 

%\ed{Demonstrate correctness and execution time trade-off by varying the MCMC convergence threshold.}

\subsection{Computational Metrics}
\label{sec:compMetrics}

The following metrics should be reported by the challenge participants to characterize the computational requirements of their implementations:

\begin{itemize}[leftmargin=0.25cm]
\item Total number of edges in the graph ($E$):  This measures the amount of data processed. 
\item Execution time: The total amount of time taken for the implementation to complete the partition, in seconds. 
\item Rate: This metric measures the throughput of the implementation, in total number of edges processed over total execution time ($E$/second). Figure \ref{fig:ratePlot} shows the preliminary results on this metric between three different implementations of the partition algorithm, when run on a HP ProLiant DL380 Gen9 with 56 cores of Intel(R) Xeon(R) processors at 2.40GHz, and 512 GB of HPE DDR4 memory at 2400 MHZ. The three implementations are: (1) C++ serial implementation and (2) C++ parallel implementation by Tiago Peixoto \cite{graph-tool_github}, and (3) Python serial implementation. The C++ implementations leverage the Boost Graph Library (BGL) extensively. Since the algorithm complexity is super-linear, the rate drops as the size of the graph increases, with a slope matching the change in rate according to the analytical complexity of the algorithm, $O(E \log^2 E)$. 

\begin{figure}[t] 
  \vspace{-1cm}
  \hspace{-0.53cm}
  \includegraphics[width=1.13\linewidth]{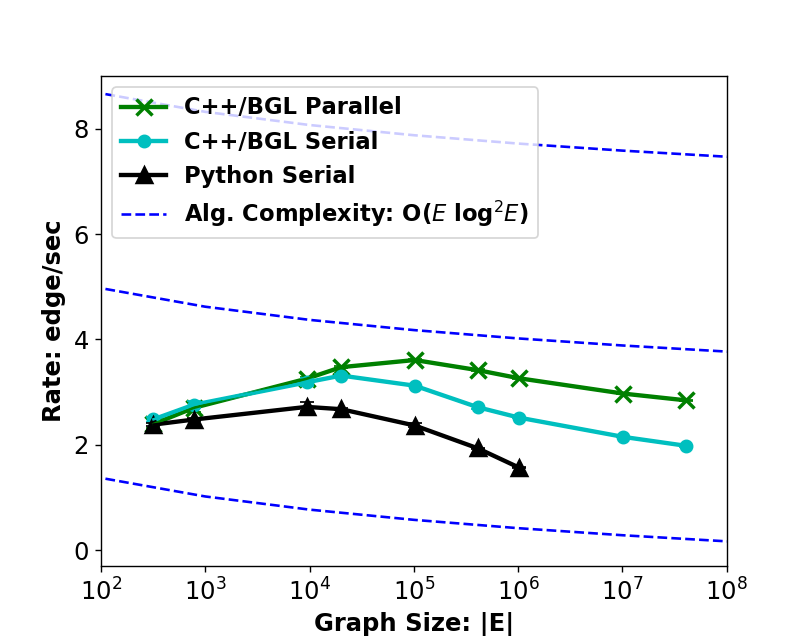}
  \caption{Processing rate for three different implementations of the baseline algorithm across graphs of increasing size. Overall, the slope of the rates follow the complexity of the algorithm, $O(E \log^2 E)$.}
  \label{fig:ratePlot} 
\end{figure}

\noindent The serial C++ implementation is about an order of magnitude faster than the Python implementation. With parallel updates, the C++ implementation gains another order of magnitude in rate when the graph is large enough. The Python implementation is limited in its ability to process very large graphs due to the lack of a fast implementation of sparse matrices in Python. All three implementations are available at \href{http://GraphChallenge.org}{GraphChallenge.org}.

\item Energy consumption in watts: The total amount of energy consumption for the computation.
\item Rate per energy: This metric captures the throughput achieved per unit of energy consumed, measured in $E$/second/Watt.
\item Memory requirement: The amount of memory required to execute the implementation.
\item Processor requirement: The number and type of processors used to execute the implementation.
\end{itemize}

\subsection{Implementation Complexity Metric}
\label{sec:compMetrics}
Total lines-of-code count: This measure the complexity of the implementation. SCLC \cite{SCLC} and CLOC \cite{CLOC} are open source line counters that can be used for this metric. The Python demonstration code for this challenge has a total of $569$ lines. The C++ open source implementation is a part of a bigger package, so it is difficult to count the lines on just the graph partition. 

\section{Summary}
This paper gives a detailed description of the graph partition challenge, its statistical foundation in the stochastic blockmodels, and comprehensive metrics to evaluate the correctness, computational requirements, and complexity of the competing algorithm implementations. This paper also recommends strategies for massively parallelizing the computation of the algorithm in order to achieve scalability for large graphs. Theoretical arguments for the correctness of the parallelization are also given. Our hope is that this challenge will provide a helpful resource to advance state-of-the-art performance and foster community collaboration in the important and challenging problem of graph partition on large graphs. Data sets and source code for the algorithm as well as metrics, with detailed documentation are available at \href{http://GraphChallenge.org}{GraphChallenge.org}.

\section{Acknowledgment}
The authors would like thank Trung Tran, Tom Salter, David Bader, Jon Berry, Paul Burkhardt, Justin Brukardt, Chris Clarke, Kris Cook, John Feo, Peter Kogge, Chris Long, Jure Leskovec, Henning Meyerhenke, Richard Murphy, Steve Pritchard, Michael Wolfe, Michael Wright, and the entire GraphBLAS.org community for their support and helpful suggestions. Also, the authors would like to recognize Ryan Soklaski, John Griffith, and Philip Tran for their help on the baseline algorithm implementation, as well as Benjamin Miller for his feedback on the matrix-based parallelism.

\bibliographystyle{unsrt}%{IEEEtran}
\bibliography{IEEEabrv,references_graph_partition}

\onecolumn
\section*{Appendix A: Partition Algorithm Pseudocode}
\label{sec:appendixA}
\vspace{-25.5cm}
\begin{algorithm*}\normalsize
\setstretch{1}
\SetKwInOut{Input}{input}\SetKwInOut{Output}{output}
\Input{$b_i^-$, $\bm{b}_{\N_i}^-$: current block labels for node $i$ and its neighbors $\N_i$ \newline
	$\bm{M}^-$: current $B \times B$ inter-block edge count matrix \newline
	$\bm{A}_{i \N_i}, \bm{A}_{\N_i i}$: edges between $i$ and all its neighbors}
\Output{$b_i^+$: the new block assignment for node $i$}
~\\

\tcp{propose a block assignment}
\textbf{obtain} the current block assignment $r= b_i^-$\\
\textbf{draw} a random edge of $i$ which connects with a neighbor $j$, obtain its block assignment $u=b_j^-$\\
\textbf{draw} a uniform random variable $x_1\sim \text{Uniform}(0,1)$\\
\eIf{$x_1 \leq \frac{B}{d^-_u+B}$} 
{
   \tcp{with some probability, propose randomly for exploration}
   \textbf{propose} $b_i^+ = s$ by drawing $s$ randomly from $\{1,2,...,B\}$ \\
}
{
   \tcp{otherwise, propose by multinomial draw from neighboring blocks to $u$}
   \textbf{propose} $b_i^+ = s$ from MultinomialDraw$\left(\frac{M^-_{u\bigcdot} + M^-_{\bigcdot u}}{d^-_u}\right) $\\
} 
\tcp{accept or reject the proposals}
\eIf{$s=r$} 
{
   \textbf{return} $b_i^+ = b_i^-$ \tcp{proposal is the same as the old assignment. done!}
}
{
   \textbf{compute} $\bm{M}^+$ under proposal (update only rows and cols $r$ and $s$, on entries for blocks connected to $i$)\\
   \textbf{compute} proposal probabilities for the Hastings correction:\\
   \hspace {0.1 in} $p_{r \rightarrow s} = \sum_{t \in \{\bm{b}_{\N_i}^-\}} \left[ K_{it} \frac{M_{ts}^- + M_{st}^- + 1}{d^-_t+B}\right]$
   \hspace {0.05 in} and \hspace {0.05 in}  $p_{s \rightarrow r} = \sum_{t \in \{\bm{b}_{\N_i}^-\}} \left[ K_{it} \frac{M_{tr}^+ + M_{rt}^+ +1}{d_t^++B}\right]$\\
   \textbf{compute} change in log posterior ($t_1$ and $t_2$ only need to cover rows and cols $r$ and $s$): \\
   \hspace {0.5 in} $\Delta S = \sum_{t_1, t_2} {\left[- M_{t_1 t_2}^+ \log\left(\frac{M_{t_1 t_2}^+}{d_{t_1,\rm out}^+ d_{t_2, \rm in}^+}\right) + M_{t_1 t_2}^- \log\left(\frac{M_{t_1 t_2}^-}{d_{t_1,\rm out}^- d_{t_2, \rm in}^-}\right)\right]}$\\
   \textbf{compute} probability of acceptance: \\
   \hspace {0.5 in} $p_{\text{accept}} = \min \left[ \exp(-\beta\Delta S) \frac{p_{s \rightarrow r}}{p_{r \rightarrow s}}, 1 \right]$\\
   \textbf{draw} a uniform random variable $x_3\sim \text{Uniform}(0,1)$\\
   \eIf{$x_3 \leq p_{\text{accept}}$}
   {
      \textbf{return} $b_i^+ = s$ \hspace {0.2 in}\tcp{accept the proposal} 
   }
   {
      \textbf{return} $b_i^+ = r$ \hspace {0.2 in}\tcp{reject the proposal} 
   }
}
\caption{Block Assignment Update At Each Node $i$}
\label{alg:NodalUpdate}
\end{algorithm*}

\pagebreak
\section*{Appendix B: Matrix-Based Batch Update Pseudocode}
\label{sec:appendixB}
\begin{algorithm*}\normalsize
\setstretch{1}
\SetKwInOut{Input}{input}\SetKwInOut{Output}{output}
\Input{$\bm{\Gamma}^-$: current block assignment matrix for all nodes \newline
	$\bm{M}^-$: current $B \times B$ inter-block edge count matrix \newline
	$\bm{A}$: graph adjacency matrix}
\Output{$\bm{\Gamma}^+$: new block assignments for all nodes}
~\\

\tcp{propose new block assignments}
\textbf{compute} node degrees: $\bm{k} = (\A + \A^T)\bm{1}$ \\
\textbf{compute} block degrees: $\bm{d}^-_{\rm out} = \bm{M}^-\bm{1}$ ; $\bm{d}^-_{ \rm in} =\bm{M}^{-^T}\bm{1}$ ; $\bm{d}^- = \bm{d}^-_{\rm out} + \bm{d}^-_{\rm in}$ \\
\textbf{compute} probability for drawing each neighbor: $\bm{P}_{\rm Nbr} = \text{RowDivide}(\A+\A^T, \bm{k} )$ \\
\textbf{draw} neighbors ($\bm{N}_{\rm br}$ is a binary selection matrix): $\bm{N}_{\rm br} = \text{MultinomialDraw}(\bm{P}_{rn})$ \\
\textbf{compute} probability of uniform random proposal: $\bm{p}_{\rm UnifProp} = \frac{B}{\bm{N}_{\rm br}\bm{\Gamma}^-\bm{d}^-+B}$ \\
\textbf{compute} probability of block transition: $\bm{P}_{\rm BlkTran} = \text{RowDivide}(\bm{M}^- + \bm{M}^{-^T}, \bm{d}^-)$ \\
\textbf{compute} probability of block transition proposal: $\bm{P}_{\rm BlkProp} = \bm{N}_{br}\bm{\Gamma}^-\bm{P}_{\rm BlkTran}$ \\
\textbf{propose} new assignments uniformly:  $\bm{\Gamma}_{\rm Unif} = \text{UniformDraw}(B, N)$ \\
\textbf{propose} new assignments from neighborhood:  \small $\bm{\Gamma}_{\rm Nbr} = \text{MultinomialDraw}(\bm{P}_{\rm BlkProp})$ \normalsize \\
\textbf{draw} $N$ Uniform$(0,1)$ random variables $\bm{x}$ \\
\textbf{compute} which proposal to use for each node: $\bm{I}_{\rm UnifProp} = \bm{x} \leq \bm{p}_{\rm UnifProp}$ \\
\textbf{select} block assignment proposal for each node:  \\
   \hspace {0.2 in}  $\bm{\Gamma}^P = \text{RowMultiply}(\bm{\Gamma}_{\rm Unif}, \bm{I}_{\rm UnifProp} ) + \text{RowMultiply}(\bm{\Gamma}_{\rm Nbr}, (\bm{1}-\bm{I}_{\rm UnifProp})) $ \\
\tcp{accept or reject the proposals}
\textbf{compute} change in edge counts by row and col: ${\Delta \bm{M}}^+_{ \rm row} = \A\bm{\Gamma}^-$ ; ${\Delta \bm{M}}^+_{\rm col} = \A^T\bm{\Gamma}^-$\\
\textbf{update} edge count matrix for each proposal: (resulting matrix is $N \times P \times P$): \\ 
   \hspace {0.2 in} $\M^+_{ijk} = M^-_{jk} - \Gamma^-_{ij} {\Delta M}^+_{{\rm row},ik} + \Gamma^P_{ij}{\Delta M}^+_{{\rm row},ik} - \Gamma^-_{ik}{\Delta M}^+_{{\rm col},ij} + \Gamma^P_{ik}{\Delta M}^+_{{\rm col},ij} $ \\
\textbf{update} block degrees for each proposal: (resulting matrix is $N \times P$): \\ 
   \hspace {0.2 in} $D^+_{{\rm out},ij} = d^-_{\rm out,j} - \Gamma^-_{ij}\sum_k{{\Delta M}^+_{{\rm row},ik}} + \Gamma^P_{ij}\sum_k{{\Delta M}^+_{{\rm row},ik}}$ \\
   \hspace {0.2 in} $D^+_{{\rm in},ij} = d^-_{{\rm in},j} - \Gamma^-_{ij}\sum_k{{\Delta M}^+_{{\rm col},ik}} + \Gamma^P_{ij}\sum_k{{\Delta M}^+_{{\rm col},ik}}$ \\
\textbf{compute} the proposal probabilities for Hastings correction ($N \times 1$ vectors): \\
   \hspace {0.2 in} $\bm{p}_{r \rightarrow s} = \left[ (\bm{p}_{\rm Nbr} \bm{\Gamma}^-) \circ (\bm{\Gamma}^P \bm{M}^- + \bm{\Gamma}^P \bm{M}^{-^T} + 1) \circ \text{RepMat}(\frac{1}{\bm{d}^- + B}, N) \right] \bm{1}$\\ 
   \hspace {0.2 in} $\bm{p}_{s \rightarrow r, i} = \left[ (\bm{p}_{\rm Nbr} \bm{\Gamma}^-) \circ (\bm{\Gamma}^- \bm{\M}^+_{i\bigcdot \bigcdot} + \bm{\Gamma}^- \bm{\M}^{+^T}_{i\bigcdot\bigcdot} + 1) \circ \frac{1}{\bm{D}_{\rm out}^+ + \bm{D}_{\rm in}^+ + B} \right] \bm{1}$\\ 
\textbf{compute} change in log posterior (only need to operate on the impacted rows and columns corresponding to $r$, $s$, and the neighboring blocks to $i$):\\
   \hspace {0.2 in} ${\Delta S}_i = \sum_{jk}{\left[-\M^+_{ijk} \log \left(\frac{\M^+_{ijk}}{D^+_{{\rm out}, ij} + D^+_{{\rm in}, ik}} \right) + M^-_{jk} \log\left(\frac{M^-_{jk}}{d^-_{{\rm out}, j} + d^-_{{\rm in}, k}}\right)\right]}$ \\

\textbf{compute} probabilities of accepting the proposal ($N \times 1$ vector):  \\
   \hspace {0.2 in} $\bm{p}_{\text{\rm Accept}} = \min \left[ \exp(-\beta {\Delta \bm{S}}) \circ \bm{p}_{s \rightarrow r} \circ \frac{1}{\bm{p}_{r \rightarrow s}}, \bm{1} \right]$\\
\textbf{draw} $N$ Uniform$(0,1)$ random variable $\bm{x}_{\rm Accept}$\\
\textbf{compute} which proposals to accept: $\bm{I}_{\rm Accept} = \bm{x}_{\rm Accept} \leq \bm{p}_{\rm Accept}$ \\
\textbf{return} $\; \bm{\Gamma}^+ = \text{RowMultiply}(\bm{\Gamma}^{P}, \bm{I}_{\rm Accept} ) + \text{RowMultiply}(\bm{\Gamma}^-, (\bm{1}-\bm{I}_{\rm Accept})) $ \\
\caption{Batch Assignment Update for All Nodes}
\label{alg:batchAssignmentUpdate}
\end{algorithm*}

\twocolumn
\section*{Appendix C: List of Notations}
\label{sec:appendixC}
\vspace{0.3cm}
\noindent Below is a list of notations used in this document:
\vspace{0.3cm}
\begin{description}[itemsep=2pt]
%\begin{itemize}[leftmargin=0.1cm, itemsep=0.1cm, align=left, labelindent=32pt]
%\begin{enumerate}[labelindent=16pt]
\item[$N$:] Number of nodes in the graph
\item[$B$:] Number of blocks in the partition 
\item[$\A$:] Adjacency matrix of size $N \times N$, where $A_{ij}$ is the edge weight from node $i$ to $j$
\item[$\bm{k}$:] Node degree vector of $N$ elements, where $k_i$ is the total (i.e.\ both in and out) degree of node $i$
\item[$\bm{K}$:] Node degree matrix of $N \times B$ elements, where $k_{it}$ is the total number of edges between node $i$ and block $t$ 
\item[$\N_i$:] Neighborhood of node $i$, which is a set containing all the neighbors of $i$ 
\item[$^-$:] Superscript that denotes any variable from the previous MCMC iteration
\item[$^+$:] Superscript that denotes any updated variable in the current MCMC iteration
\item[$\bm{b}$:] Block assignment vector of $N$ elements where $b_i$ is the block assignment for node $i$ 
\item[$\bm{\Gamma}$:] Block assignment matrix of $N \times B$ elements where each row $\bm{\Gamma}_{i\bigcdot}$ is a binary indicator vector with $1$ only at the block node $i$ is assigned to. $\bm{\Gamma}^P$ is the proposed block assignment matrix.
\item[$\bm{M}$:] Inter-block edge count matrix of size $B \times B$, where $M_{ij}$ is the number of edges from block $i$ to $j$
\item[$\bm{\M}^+$:] Updated inter-block edge count matrix for each proposal, of size $N \times B \times B$
\item[${\Delta \bm{M}}^+_{ \rm row/col}$:] Row and column updates to the inter-block edge count matrix, for each proposal. This matrix is of size $N \times B$.
\item[$\bm{d}_{ \rm in}$:] In-degree vector of $B$ elements, where $d_{ \rm in,i}$ is the number of edges into block $i$
\item[$\bm{d}_{\rm out}$:] Out-degree count vector of $B$ elements, where $d_{\rm out,i}$ is the number of edges out of block $i$
\item[$\bm{d}$:] Total edge count vector of $B$ elements, where $d_i$ is the total number of edges into and out of block $i$. $\bm{d} = \bm{d}_{ \rm in} + \bm{d}_{\rm out}$
\item[$\bm{D}^+_{\rm in/out}$:] In and out edge count matrix for each block, on each proposal. It is of size $N \times B$ 
\item[$\Delta S$:] The difference in log posterior between the previous block assignment and the new proposed assignment
\item[$\beta$:] Learning rate of the MCMC
\item[$p_{r \rightarrow s}$:] Probability of proposing block $s$ on the node to be updated which currently is in block $r$
\item[$p_{\rm Accept}$:] Probability of accepting the proposed block on the node
\item[$\bm{P}_{\rm Nbr}$:] Matrix of $N \times N$ elements where each element $\bm{P}_{Nbr, ij}$ is the probability of selecting node $j$ when updating node $i$
\item[$\bm{N}_{\rm br}$:] Matrix of $N \times N$ elements where each row $\bm{N}_{{\rm br}, i\bigcdot}$ is a binary indicator vector with $1$ only at $j$, indicating that $j$ is selected when updating $i$
\item[$\bm{p}_{\rm UnifProp}$:] Vector of $N$ elements representing the probability of uniform proposal when updating each node
\item[$\bm{P}_{\rm BlkTran}$:] Matrix of $B \times B$ elements where each element $\bm{P}_{{\rm BlkTran}, ij}$ is the probability of landing in block $j$ when randomly traversing an edge from block $i$
\item[$\bm{P}_{\rm BlkProp}$:] Matrix of $N \times B$ elements where each element $\bm{P}_{{\rm BlkProp}, ij}$ is the probability of proposing block assignment $j$ for node $i$
\item[$\bm{\Gamma}_{\rm Unif}$:] Block assignment matrix from uniform proposal across all blocks. It has $N \times B$ elements where each row $\bm{\Gamma}_{{\rm Unif}, i\bigcdot}$ is a binary indicator vector with $1$ only at the block node $i$ is assigned to
\item[$\bm{\Gamma}_{\rm Nbr}$:] Block assignment matrix from neighborhood proposal. It has $N \times B$ elements where each row $\bm{\Gamma}_{{\rm Unif}, i\bigcdot}$ is a binary indicator vector with $1$ only at the block node $i$ is assigned to
\item[$\bm{I}_{\rm UnifProp}$:] Binary vector of $N$ elements with $1$ at each node taking the uniform proposal and $0$ at each node taking the neighborhood proposal
\item[$\bm{I}_{\rm Accept}$:] Binary vector of $N$ elements with $1$ at each node where the proposal is accepted and 0 where the proposal is rejected
\item[${\rm Uniform}(x,y)$:] Uniform distribution with range from $x$ to $y$ 
\item[$\delta_{tk}$:] Dirac delta function which equals $1$ if $t=k$ and $0$ otherwise.
\item[${\rm RowDivide}(\bm{A},\bm{b})$:] Matrix operator that divides each row of matrix $\bm{A}$ by the corresponding element in vector $\bm{b}$ 
\item[${\rm RowMultiply}(\bm{A},\bm{b})$:] Matrix operator that multiplies each row of matrix $\bm{A}$ by the corresponding element in vector $\bm{b}$ 
\item[${\rm UniformDraw}(B, N)$:] Uniformly choose an element from $\{1,2,...,B\}$ as the block assignment $N$ times for each node, and return a $N \times B$ matrix where each row $i$ is a binary indicator vector with $1$ only at $j$, indicating node $i$ is assigned block $j$ 
\item[${\rm MutinomialDraw}(\bm{P}_{\rm BlkProp})$:] For each row of the proposal probability matrix $\bm{P}_{{\rm BlkProp}, i\bigcdot}$, draw an block according to the multinomial probability vector $\bm{P}_{\rm{BlkProp}, i\bigcdot}$ and return a $N \times B$ matrix where each row $i$ is a binary indicator vector with $1$ only at $j$, indicating node $i$ is assigned block $j$ 
%\end{itemize}
%\end{enumerate}
\end{description}

\end{document}